\documentclass{emulateapj}
\submitted{{\it Submitted for publication in ApJ Letters}}
\usepackage{multirow,color,wrapfig,ulem}
\usepackage {graphicx}
\usepackage{graphics}
\usepackage[dvips]{epsfig}

\newcommand{\kms}{\,km~s$^{-1}$}

\newcommand{\beq}{\begin{eqnarray}}  
\newcommand{\eeq}{\end{eqnarray}}

\newcommand{\ly}{{\ifmmode{{\rm Ly}\alpha}\else{Ly$\alpha$}\fi}}
\newcommand{\hMpc}{{\ifmmode{h^{-1}{\rm Mpc}}\else{$h^{-1}$Mpc }\fi}}  
\newcommand{\hGpc}{{\ifmmode{h^{-1}{\rm Gpc}}\else{$h^{-1}$Gpc }\fi}}  
\newcommand{\hmpc}{{\ifmmode{h^{-1}{\rm Mpc}}\else{$h^{-1}$Mpc }\fi}}  
\newcommand{\hkpc}{{\ifmmode{h^{-1}{\rm kpc}}\else{$h^{-1}$kpc }\fi}}  
\newcommand{\hMsun}{{\ifmmode{h^{-1}{\rm {M_{\odot}}}}\else{$h^{-1}{\rm{M_{\odot}}}$}\fi}}  
\newcommand{\hmsun}{{\ifmmode{h^{-1}{\rm {M_{\odot}}}}\else{$h^{-1}{\rm{M_{\odot}}}$}\fi}}  
\newcommand{\Msun}{{\ifmmode{{\rm {M_{\odot}}}}\else{${\rm{M_{\odot}}}$}\fi}}  
\newcommand{\msun}{{\ifmmode{{\rm {M_{\odot}}}}\else{${\rm{M_{\odot}}}$}\fi}}

\newcommand{\rand}{{\ifmmode{{\mathcal{R}}}\else{${\mathcal{R}}$ }\fi}}

\def\spose#1{\hbox to 0pt{#1\hss}}
\def\simlt{\mathrel{\spose{\lower 3pt\hbox{$\mathchar"218$}}
     \raise 2.0pt\hbox{$\mathchar"13C$}}}
\def\simgt{\mathrel{\spose{\lower 3pt\hbox{$\mathchar"218$}}
     \raise 2.0pt\hbox{$\mathchar"13E$}}}

\shorttitle{Local Group Kinematics}
\shortauthors{Forero-Romero et al.}

\begin{document}
\title{The kinematics of the Local Group in a cosmological context}
\author{
J. E.\ Forero-Romero\altaffilmark{1}, 
Y. Hoffman\altaffilmark{2}, 
S. Bustamante\altaffilmark{3}, 
S. Gottl\"ober\altaffilmark{4}, 
G. Yepes\altaffilmark{5}
}

\altaffiltext{1}{Departamento de F\'{i}sica, Universidad de los Andes, Cra. 1 No. 18A-10, Edificio Ip, Bogot\'a, Colombia, \email{je.forero@uniandes.edu.co}}
\altaffiltext{2}{Racah Institute of Physics, The Hebrew University of Jerusalem, 91904 Jerusalem, Israel}
\altaffiltext{3}{Instituto de F\'{\i}sica - FCEN, Universidad de Antioquia, Calle 67 No. 53-108, Medell\'{\i}n, Colombia}
\altaffiltext{4}{Leibniz-Institut f\"ur Astrophysik, Potsdam, An der Sternwarte 16, 14482 Potsdam, Germany}
\altaffiltext{5}{Grupo de Astrof\'{\i}sica, Departamento de F\'{\i}sica Te\'orica, Universidad Aut\'onoma de Madrid,
Cantoblanco E-280049, Spain}

\date{\today}

\begin{abstract}
Recent observations constrained the tangential velocity of M31 with
respect to the Milky Way (MW) to be $v_{\rm M31,tan}<34.4$ \kms and
the radial velocity to be in the range $v_{\rm M31,rad}=-109\pm
4.4$\kms \citep{vanderMarel12}. In this study we use a large volume
high resolution N-body cosmological simulation (Bolshoi) together with three
constrained simulations to statistically study this kinematics in the
context of the $\Lambda$CDM. The comparison of the ensembles of
simulated pairs with the observed LG at the 1-$\sigma$ level in
  the uncertainties has been done with respect to the
radial and tangential velocities, the reduced orbital energy
($e_{\rm tot}$), angular momentum ($l_{\rm orb}$) and the
dimensionless spin parameter, $\lambda$. Our main results are: (i) the
preferred radial and tangential velocities for pairs in $\Lambda$CDM
are $v_{\rm r}=-80\pm 20$\kms, $v_{\rm   t}=50\pm 10$\kms, (ii) pairs
around that region are $3$ to $13$ times more common than pairs within
the observational values, (iii) $15\%$ to $24\%$ of LG-like pairs in
$\Lambda$CDM have energy and angular momentum consistent with
observations while (iv) $9\%$ to $13\%$ of pairs in the same sample
show similar values in the inferred dimensionless spin parameter. It
follows that within current observational uncertainties the
quasi-conserved quantities that characterize the orbit of
the LG, i.e. $e_{\rm tot}$, $l_{\rm   orb}$ and $\lambda$, do not
challenge the standard $\Lambda$CDM model, but the model is in tension
with regard to the actual values of the radial and tangential
velocities. This might hint to a problem of the $\Lambda$CDM model to
reproduce the observed LG.  

\end{abstract}

\begin{keywords}
{galaxies: kinematics and dynamics, Local Group, methods:numerical}
\end{keywords}

\section{Introduction}

The Milky Way (MW) and Andromeda galaxy (M31) are the dominant
galaxies in the Local Group (LG). Astronomical observations of their
mass distribution impose constraints on the standard cosmological
model. The satellite overabundance problem \citep{Klypin99,Moore99},
tidal disruption features \citep{pandas09} and the disk dominated
morphology \citep{Kazantzidis2008} are examples of how LG studies
are linked to the cosmological context. Detailed studies on the
Magellanic Clouds dynamics and their possible link to M31 add to the
interest of understanding the details of the LG kinematics and
dynamics
\citep{Besla2007,Tollerud2011,Knebe2011,Fouquet2012,Teyssier2012}. However,
a general concern in the use of the LG as a tool for near-field
cosmology \citep{Freeman2002,Peebles2010} is how typical is the LG
regarding the properties of interest
\citep{Busha2011,Liu2011,ForeroRomero2011,Purcell2012}.  

A new valuable piece of information in this issue is the recent
observational determination of the proper-motion measurements of M31,
which until recently had been out of reach \citep{vanderMarel12}.  The
reported measurements set an upper bound for the tangential velocity
of M31 with respect to the MW of $v_{\rm tan,M31}\leq 34.4$ km
s$^{-1}$. Together with the values of the relative radial velocity of
$v_{\rm rad,M31}=-109\pm 4.4$ \kms observations show that the relative
motion of the MW and Andromeda is consistent with a head-on
collision. With this information it is possible to quantify how common
such a kinematic configuration is in a $\Lambda$CDM Universe. 

This Letter presents such a study. We use a large volume, high resolution
dark matter only N-body simulation in the concordance $\Lambda$CDM
cosmology to find a set of halo pairs with similar  characteristics as
inferred in the LG. We quantify these results in terms of the number
of pairs with given radial and tangential velocities in the
galactocentric rest frame. We also find the pairs that are consistent
with a head on collision in terms of the ratio of the radial to
tangential velocity $f_{\rm t}\equiv v_{\rm tan}/v_{\rm rad}<0.32$ and
present these results in terms of the reduced angular momentum,
mechanical energy and dimensionless spin parameter.

In addition we make use of three  constrained N-body simulations which
are constructed to reproduce the observed large scale structure of the
Local Universe on scales of a few tens of Mpc. The special feature of
these simulations is that each volume features a pair of halos with
the right characteristics to be considered LG-like objects. 

This Letter is structured as follows. In the next section we present
the N-body simulations and the criteria we use to select LG-like halo
pairs. In Section \ref{sec:results} we present the results for the
dynamics in these pairs in terms of the tangential/radial velocities
and the orbital angular momentum/mechanical energy. In the same
section we summarize these dynamical results in terms of the
dimensionless spin parameter of the pairs. Finally, in the last
section we comment and conclude about the implications of these
results in the context of the $\Lambda$CDM model.

\section{Simulation and Pair Samples}
\label{sec:methods}
\subsection{The Bolshoi and Constrained Simulations}

The Bolshoi simulation follows the non-linear evolution of the dark
matter density field using N-body techniques. The simulation has a
cubic comoving volume of $(250\hMpc )^3$, sampled with $2048^{3}$
particles. The cosmological parameters used in the simulation are
$\Omega_{m}=0.27, \Omega_{\Lambda}=0.73, \sigma_{8}=0.82, h=0.70$ and
$n=0.95$, corresponding to the matter density, vacuum energy density,
the normalization of the power spectrum, the dimensionless Hubble
constant and the index of the slope in the initial power
spectrum. This set of parameters is compatible with the analysis  of
the seventh year of data from the Wilkinson Microwave Anisotropy Probe
(WMAP) \citep{Jarosik2011}. A detailed description of this simulation
can be found in \cite{Bolshoi}. 

With these parameters the mass per particle is $m_{p}=1.4\times
10^{8}$\hMsun. In this paper we use halos obtained through the
Bound Density Maxima (BDM) algorithm \citep{KlypinBDM}. The halos are
selected to have an overdensity of 200 times the critical
density. Furthermore, we only include in the analysis halos whose
center is located outside the virial radius of any other halo.  We
have obtained the data through the public available Multidark
database \footnote{{\tt http://www.multidark.org/MultiDark/}}
\citep{2011arXiv1109.0003R}. The database allows us to obtain the
comoving positions, peculiar velocities and masses for all the halos
in the simulation volume at $z=0$. The positions and velocities of
these haloes correspond to the average values of the $250$ most bound
particles. The Hubble flow is taken into account to convert the
peculiar velocities into physical velocities and allow for a
comparison with observations. We have verified that the main
conclusions of this paper hold in the case of halos defined by a FOF
algorithm with a linking length 0.17 times the mean interparticle
distance. 

The constrained simulations we use in this Letter are part of the
Constrained Local UniversE Simulations (CLUES)  \footnote{{\tt http://www.clues-project.org}} project whose main
objective is to reproduce the large scale structure in the Local
Universe as accurately as possible. The algorithm and observational
constraints to construct the initial conditions are  described in
\cite{clues2010}.  We use three dark matter only simulations, each
has a cubic volume of $64$\hMpc on a side, with the density field
sampled with $1024^3$ particles. The cosmological density parameter is
$\Omega_m=0.28$, the cosmological constant $\Omega_{\Lambda}=0.72$,
the dimensionless Hubble parameter $h=0.73$, the spectral index of the
primordial density perturbations $n=0.96$ and the power spectrum
normalization $\sigma_{8}=0.817$, also consistent with WMAP 7th year
data.

\subsection{Two samples of LG-like pairs}
Based on the BDM catalogs in the Bolshoi simulation we construct a
halo pair sample with the dynamical  properties consistent with  those
of  the MW and M31. The  criteria  we impose to define a LG-like halo
pair are the following: 

\begin{enumerate}
\item Each halo has a mass in the range $7\times10^{11}\Msun
  <M_{h}<7\times 10^{12}\Msun$. 
\item With respect to each halo, there cannot be any other halo within
  the mass range $7\times10^{11}\Msun <M_{h}<7\times 10^{12}\Msun$
  closer than its partner. It means that there cannot be ambiguity on
  the identity of the pair members. 
\item The relative radial velocity between the two halos is negative
  \citep{vanderMarel12}. 
\item The distance between the center of mass of the halos must be
  less than $1.0$Mpc \citep{ribas05,vanderMarel08}. 
\item There cannot be halos more massive than $7\times 10^{12}$\Msun
  within a radius of $3$Mpc with respect to every object centre
  \citep{Karachentsev04,Anton09}. 
\item There cannot be halos more massive than $7\times 10^{13}$\Msun
  within  a radius of $4$Mpc with respect to every object centre
  \citep{Karachentsev04}. 
\end{enumerate}

Throughout this Letter we refer to this sample as the full
sample. 
This sample in the Bolshoi simulation has $1923$ pairs. Additionally,
there is a sample of three (3) pairs constructed from the three
constrained realizations. These pairs fulfill all the above mentioned
conditions and additionally are located in a place with the right
distances with respect to the Virgo cluster in the simulation.

The full observational characteristics that we take in this Letter for
the MW-M31 pair are listed in Table \ref{table:1}.  
A more reduced sample from the full sample has been constructed
so as to obey the observational bounds on the masses and
separation of the two main halos. These amount to:   
\begin{enumerate}
\item The separation between the center of mass of the halos is in the
  range $700-800$kpc \citep{ribas05,vanderMarel08}. 
\item The total mass of the two halos is in the range $1-4\times
  10^{12}$\Msun \citep{vanderMarel12}. 
\end{enumerate}

Including these conditions the full sample is reduced from $1923$
to $158$ pairs. We refer to this sample as the reduced sample. Note
that only one constrained LG-like object is included in this subset.

\begin{table}
\caption{Summary of kinematic observational constraints.}
\begin{center}
\begin{tabular}{ccc}\hline
$v_{\rm M31,rad}$ &(\kms) & $-109.3\pm 4.4$\\
$v_{\rm M31,tan}$ &(\kms) & $<34.4$\\
$r_{\rm M31}$ &(kpc) & $770\pm 40$\\
${\bf r}_{\rm M31}$ & (kpc) &$(-378.9, 612.7, -283.1)$\\
$\sigma_{{\bf r}, {\rm M31}}$ & (kpc) &$(-18.9, 30.6, 14.5)$\\
${\bf v}_{\rm M31}$ & (\kms) & $(66.1, -76.3, 45.1)$\\
$\sigma_{{\bf v},{\rm M31}}$ & (\kms) &$(26.7, 19.0, 26.5)$\\
$M_{\rm 200, MW}$ & ($10^{12}\Msun$) & $1.6\pm0.5$ \\
$M_{\rm 200, M31}$ & ($10^{12}\Msun$) & $1.6\pm0.5$ \\
$M_{\rm 200,MW} + M_{\rm 200, M31}$ & ($10^{12}\Msun$) & $3.14\pm 0.58$\\
$\log_{10}\lambda$& & $-1.72\pm 0.07$ \\\hline
\end{tabular}\\
\end{center}
\vspace{1mm}
Notes:
\begin{enumerate}
\item The kinematic
  properties for M31 are reported in the galactocentric restframe 
  \citep{vanderMarel12}. 

\item Values in parenthesis correspond to vector
  components. $\sigma_{\bf x}$ represents the uncertainty on the
  components of vector ${\bf x}$. The uncertainties correspond to
    1-$\sigma$ values.
\item The values for the individual halo
masses are consistent with the priors used by \cite{vanderMarel12}.

\item The observational uncertainties in the position vector
correspond to a $5\%$ in each component consistent with the
1-$\sigma$ uncertainties in the distance \citep[see references
  in][]{vanderMarel08}.

\item The value for $\log_{10}\lambda$ is obtained in this Letter
from a Monte Carlo simulation as described in Section 3.
\end{enumerate}
\label{table:1}
\end{table}

\begin{table}
\caption{Summary of results from the $\Lambda$CDM Bolshoi simulation}
\begin{center}
\begin{tabular}{cccc}\hline
 &  & Full Sample & Reduced Sample\\
$v_{\rm M31,rad}$ &(\kms) & $-70\pm 10$ & $-90\pm 10$\\
$v_{\rm M31,tan}$ &(\kms) & $50\pm 10$ & $50\pm 10$\\
$\log_{10}\lambda$& & $-1.47\pm 0.13$& $-1.34\pm 0.12$\\\hline
\end{tabular}
\end{center}
\vspace{1mm}
Note: the velocity uncertainties correspond to the minimum bin size required to
obtain robust statistics in Figure \ref{fig:rt}
\label{table:2}
\end{table}

\section{Results}
\label{sec:results}

\begin{figure*}
\begin{center}
\includegraphics[keepaspectratio=true,width=0.42\textwidth]{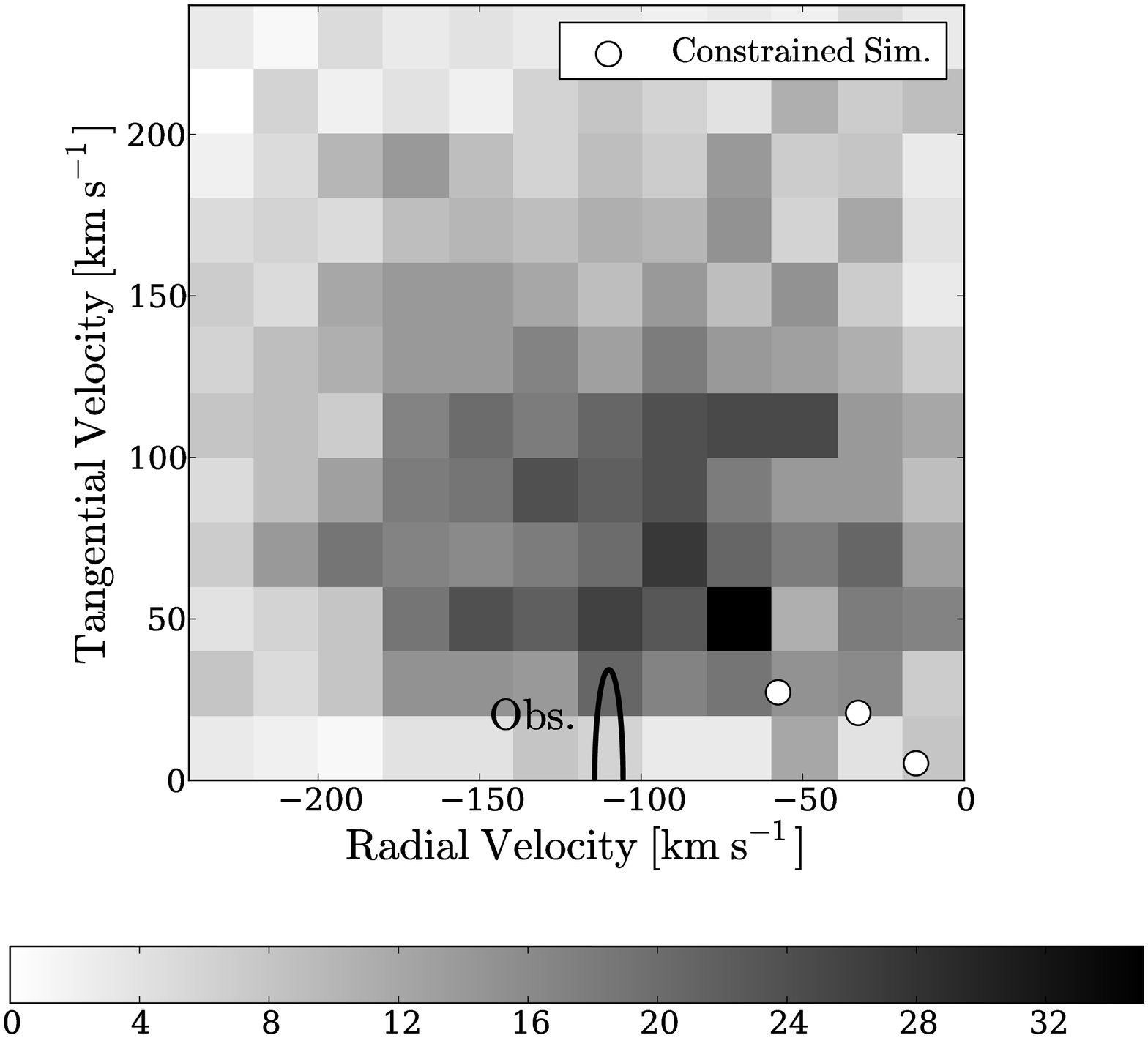}
\includegraphics[keepaspectratio=true,width=0.42\textwidth]{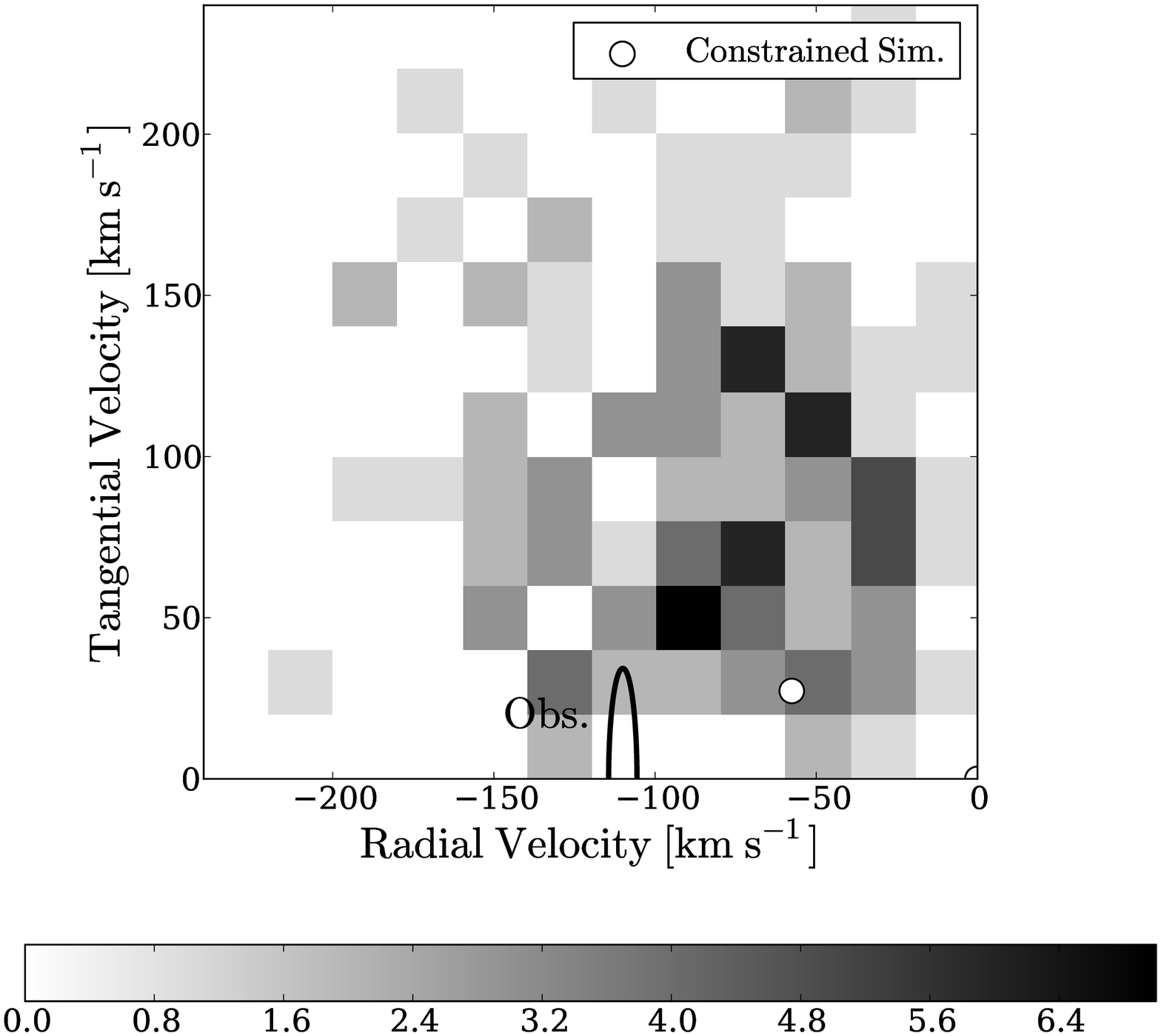}
\includegraphics[keepaspectratio=true,width=0.42\textwidth]{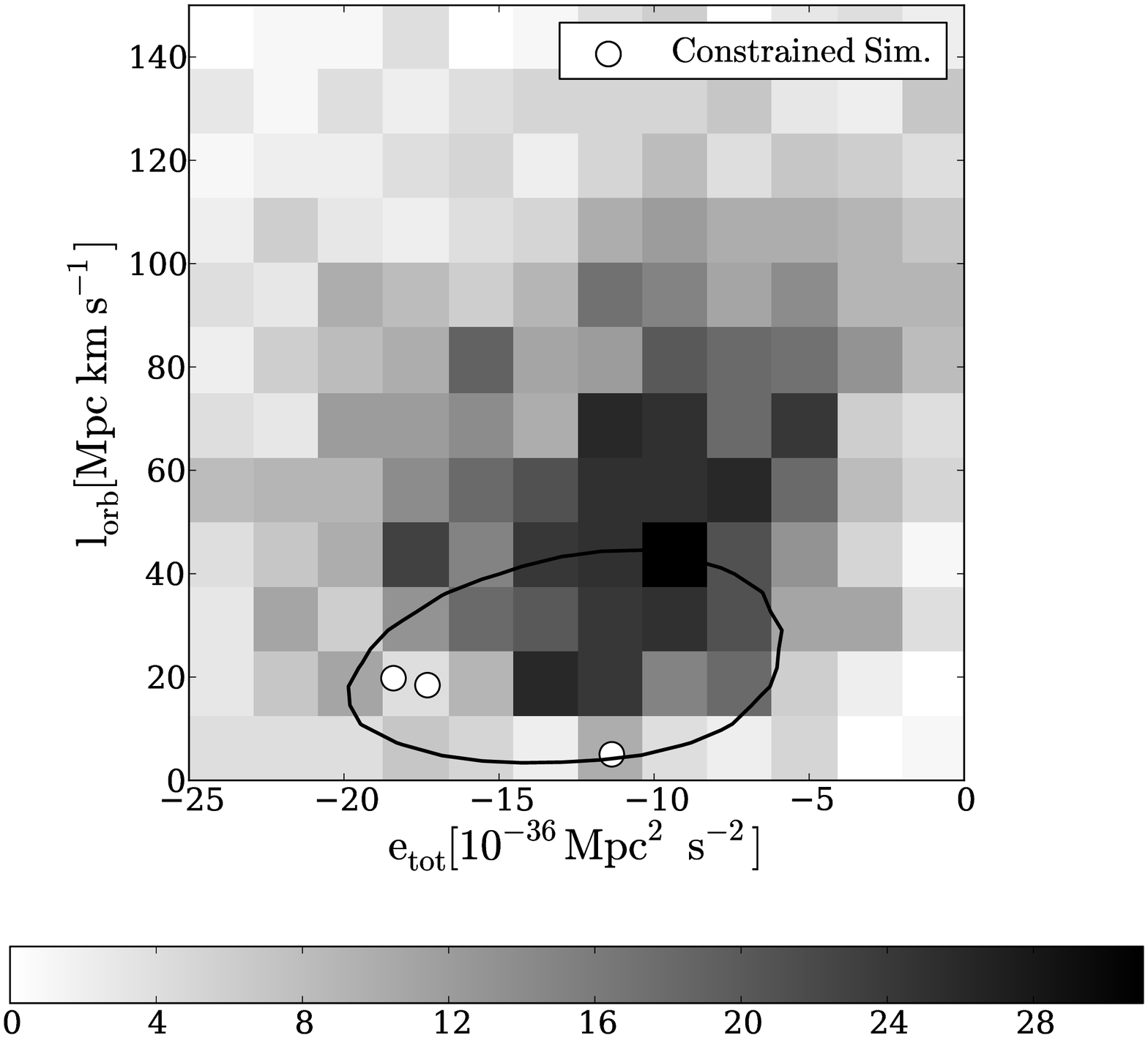}
\includegraphics[keepaspectratio=true,width=0.42\textwidth]{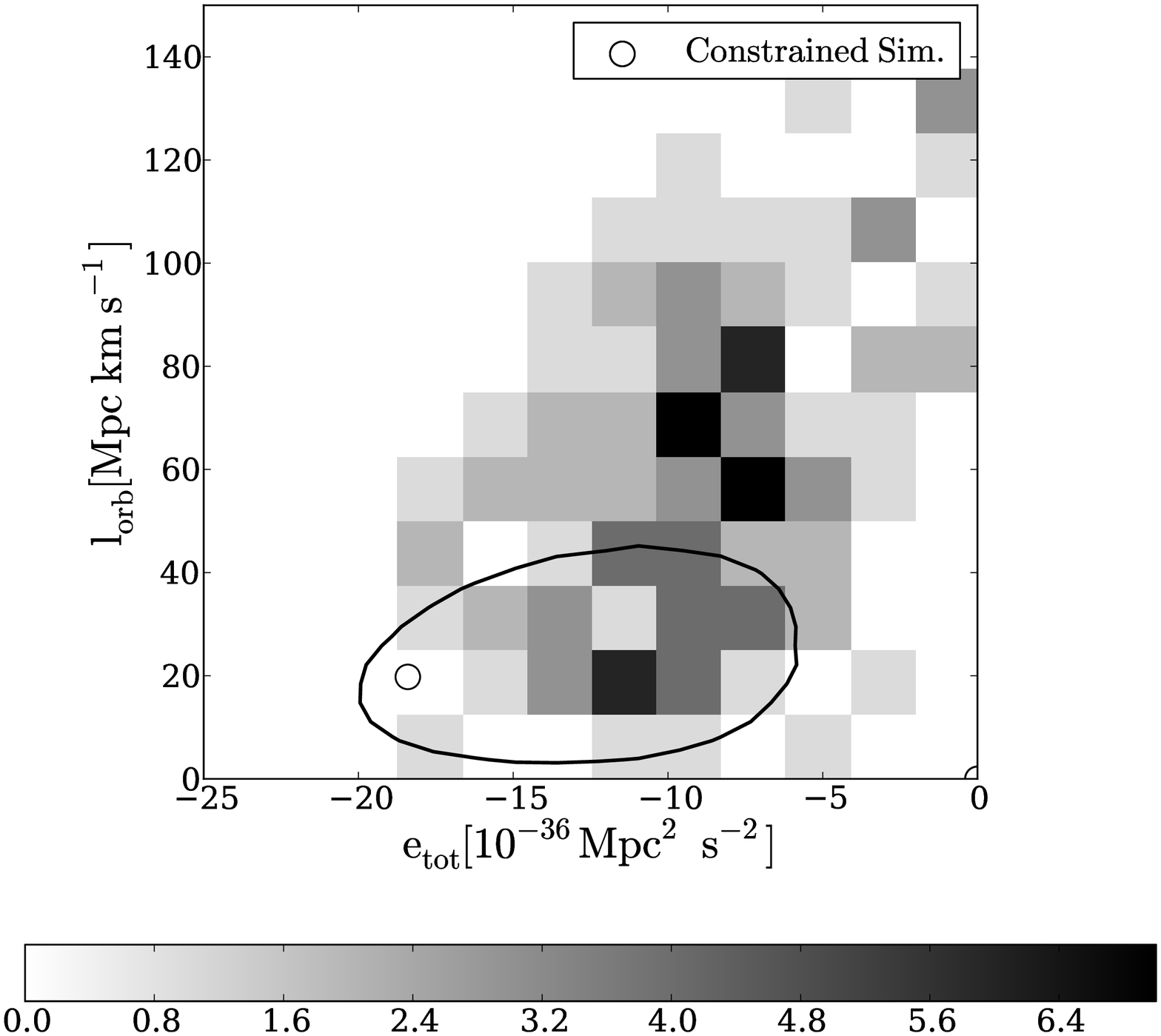}
\caption{Upper panels: 2D histograms of the radial and tangential
  velocities for LG-like halo pairs in the Bolshoi simulation.   Lower
  panels: 2D histograms of the orbital angular momentum ($l_{\rm orb}$)
  and mechanical energy ($e_{\rm tot}$) per unit of reduced mass
  calculated considering the halos as point masses. Left (right) panels
    correspond to the full (reduced) sample. The bar indicates
  the number of pairs in each cell. The locii for the high
  density regions in the $v_{\rm r}-v_{\rm t}$ plane are listed in
  Table \ref{table:2}.   
  The half ellipse in the $v_{\rm r}-v_{\rm t}$ plane corresponds to
  the 1-$\sigma$ uncertainties in observations. Equivalently, the contour
  line in   the $e_{\rm tot}-l_{\rm orb}$ plane encloses $68\%$ of the
  Monte   Carlo generated points from
  the   observational values summarized in Table 1. The circles
  represent the positions of the pairs from the constrained simulations.
}
\label{fig:rt}
\end{center}

\end{figure*}

\begin{table*}
\caption{Summary of the comparison of the observational results
  against $\Lambda$CDM.}
\begin{center}
\begin{tabular}{ccccc}\hline
Physical & (\%) Pairs consistent & (\%) Pairs consistent & (\%) Pairs
with highest & (\%) with highest\\ 
property & with observations (1-$\sigma$) & with
observations (1-$\sigma$)& likelihood in $\Lambda$CDM & likelihood in $\Lambda$CDM\\ 
 & (full sample) & (reduced sample) & (full sample) & (reduced sample)\\ \hline
$v_{\rm r}$-$v_{\rm t}$ & (0.4\%) 8/1923 & ($<$0.6\%)0/158 & (1\%)23/1923 & (8\%)13/158\\
$e_{\rm tot}$-$l_{\rm orb}$ & (15\%)298/1923 & (24\%)38/158 & - & -\\
$\log_{10}\lambda$ & (13\%)257/1923 & (9\%)15/158 & - & -\\
$r_{\rm t}=v_{\rm t}/v_{\rm r}$& (12\%)242/1923 & (8\%)13/158 & -& -\\\hline
\end{tabular}
\end{center}
\vspace{1mm}
Notes:\\
\begin{enumerate}
\item 
In columns 2 and 3 consistency is defined from 1-$\sigma$
  uncertainties for $v_{\rm r}$-$v_{\rm t}$, $e_{\rm tot}$-$l_{\rm
    orb}$ and $\log_{10}\lambda$. In the case of $r_{\rm t}$, is
  defined from the constrain $r_{\rm t}<0.32$ derived from the
  1$-\sigma$ observational uncertainties in the radial and tangential
  velocities.
\item
In columns 4 and 5 the number of pairs around the preferred region
in $v_{\rm r}-v_{\rm t}$ in $\Lambda$CDM (as shown in
Fig. \ref{fig:rt} and summarized in Table \ref{table:2}) are
calculated using the same observational 1-$\sigma$ absolute
uncertainty on $v_{\rm r}$ and $v_{\rm t}$.
\end{enumerate}
\label{table:3}
\end{table*}

\subsection{Radial and tangential velocities}

Figure \ref{fig:rt} summarizes the central finding of this
Letter. Most of the pairs in the two samples constructed from the
Bolshoi Simulation have radial and tangential velocities notably
different from the observational constraints at the 1-$\sigma$ level.    
 
The most probable radial and tangential velocities in $\Lambda$CDM are
summarized in Table \ref{table:2}. For the full sample we have $v_{\rm
  rad,\Lambda CDM} = -70\pm10$\kms and $v_{\rm tan,\Lambda CDM} =
50\pm20$ \kms. For the reduced sample $v_{\rm
  rad,\Lambda CDM} = -90\pm10$\kms and $v_{\rm tan,\Lambda CDM} =
50\pm20$ \kms, where the uncertainties in these values reflect
the minimum grid size needed to obtain robust statistics for the 2D
histogram. 

The number of pairs compatible with LG observations  at the
  1-$\sigma$ level are listed in
Table \ref{table:3} (columns 2 and 3). In the same table (columns 4
and 5) we summarize the number of pairs around the $\Lambda$CDM values
within the same range of absolute observational uncertainty 
(i.e. $\sigma_{\rm tan}=17$\kms and $\sigma_{\rm rad}=4$\kms.) 

From these results we infer that the pairs around the preferred phase
space for $\Lambda$CDM are at least $13$ times more common than pairs
with the observed velocities for the LG. We highlight that
this is a lower bound given that in the reduced sample none of the
pairs is found in the interval allowed by observations. The high
eccentric orbit of the observed LG  constitutes an unlikely
configuration for the $\Lambda$CDM LG-like objects. This holds for
the full and the reduced sample of objects. These
  conclusions are valid at 1-$\sigma$ level in the observational
  uncertainties. An equivalence in the abundance between the pairs
  around the observational values and those around the
  preferred $\Lambda$CDM velocities is reached only at
  the 4-$\sigma$ level on the observational uncertainties.

From Figure \ref{fig:rt} it is also clear that there is a significant
number of pairs with a high tangential-to-radial velocity ratio. The
peak in the pair number density is located around a region of $f_{\rm
  t}\equiv v_{\rm tan \Lambda CDM}/v_{\rm rad \Lambda CDM}\sim 0.7$,
while the observations suggest $f_{\rm t}<0.32$. As summarized in Table
\ref{table:3} we find that only between $8\%$ to $12\%$ of the pairs are
consistent with the observational constraint. The three
pairs from the constrained realizations are also show higher $f_{\rm
  t}$ ratios than observations: $f_{\rm t}= 0.35, 0.45, 0.73$.

\subsection{Reduced Angular Momentum and Energy}
\label{subsection:e-l}

The two-body problem of point-like masses can serve as a proxy for the
dynamics of the LG and can be used as a tool for  studying the LG within
the framework of the standard cosmological model. Within the model the
dynamics of the LG is governed by the (center of mass)  reduced
angular momentum $l_{\rm orb}= |{{\bf r}_{\rm M31}}\times{\bf v}_{\rm M31}|$ and
reduced energy  $e_{\rm tot}=\frac{1}{2} {\bf v}_{\rm M31}|^2 - G M/|{\bf
  r}_{\rm M31}|$,  where the total mass is $M=m_{\rm M31}+m_{\rm MW}$,
the reduced mass is $\mu=m_{\rm M31}m_{\rm MW}/M$ and $G$ is the
gravitational constant.

This formulation has a clear theoretical advantage if one considers
the angular momentum and the mechanical energy as quasi-conserved
dynamical quantities. This means that, after some formation time,
these quantities do not significantly evolve.

However, this formulation has an observational disadvantage. The
reduced energy and angular momentum are derived from very different
kinds of observations, increasing the uncertainties in their final
determination.  

We use Monte Carlo sampling to estimate the reduced energy and angular
momentum from the observed properties in the LG listed in
Table\ref{table:1}. Lower panels in Figure \ref{fig:rt} also presents the results in the
plane $e_{\rm tot}-l_{\rm orb}$\footnote{The  1-$\sigma$ contour in the lower
  panels of Figure \ref{fig:rt}
does not cross the zero level, as can be naively expected from Figure
\ref{fig:rt}. This happens for the $\approx 80\%$ contour level.}.  As
expected, the $e_{\rm tot}$  and $l_{\rm orb}$ 
constraints are less restrictive compared with the radial and
tangential velocity constraints; 15\% (24\%) of pairs in the full
(reduced) sample obey the 1-$\sigma$ observational constraints.
We note that for the reduced sample the preferred region is outside the
1-$\sigma$ observational contours.

All the constrained LGs of the full and reduced sample are
consistent with the 1-$\sigma$ observational uncertainty.  However a
an hypothetical increase by a factor of $2$ in accuracy of the
tangential velocity, showing that it is below $17$\kms, would bring
these pairs and the $\Lambda$CDM expectation outside this uncertainty
region.

\begin{figure*}
\begin{center}
\includegraphics[keepaspectratio=true,width=0.46\textwidth]{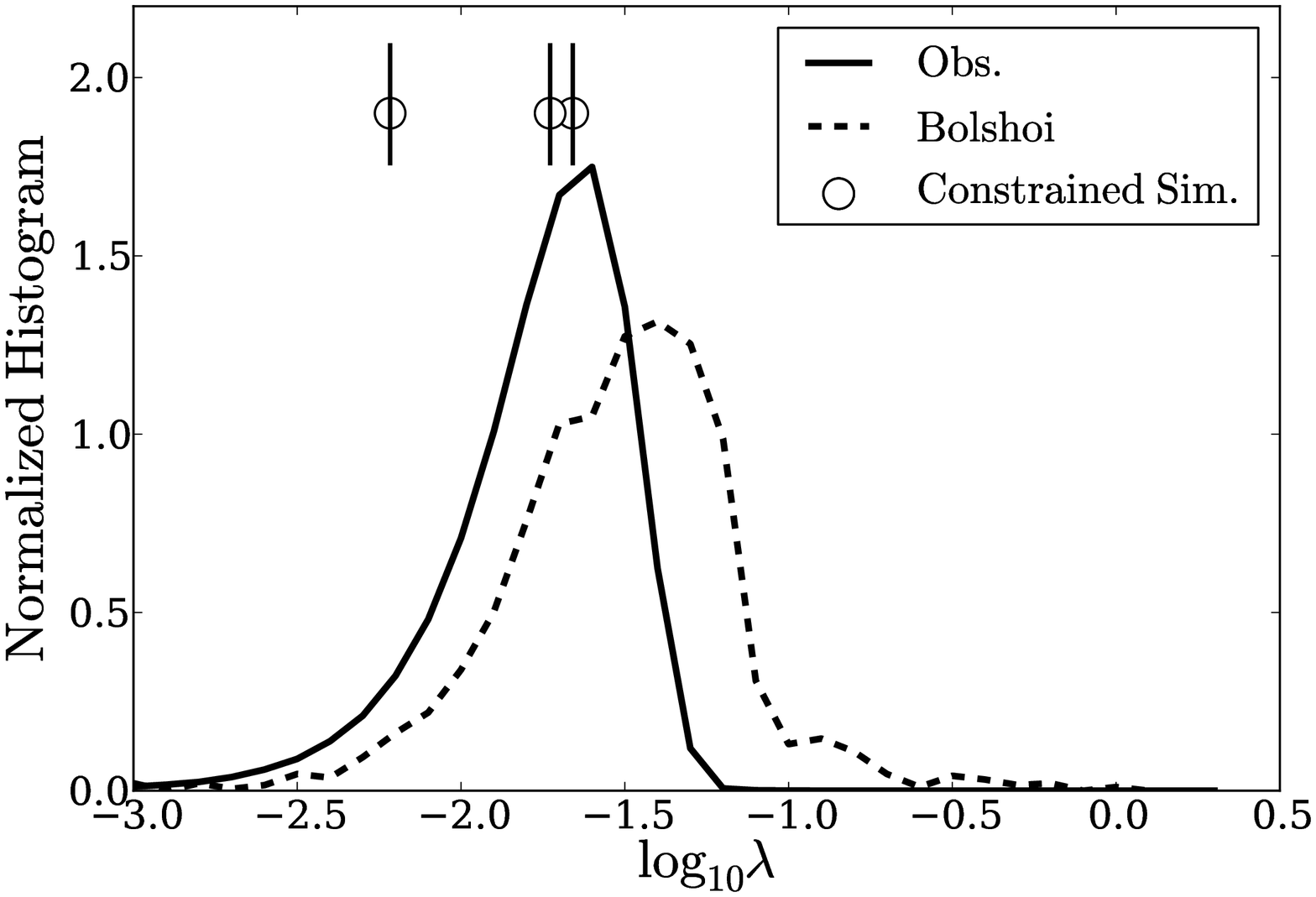}
\includegraphics[keepaspectratio=true,width=0.46\textwidth]{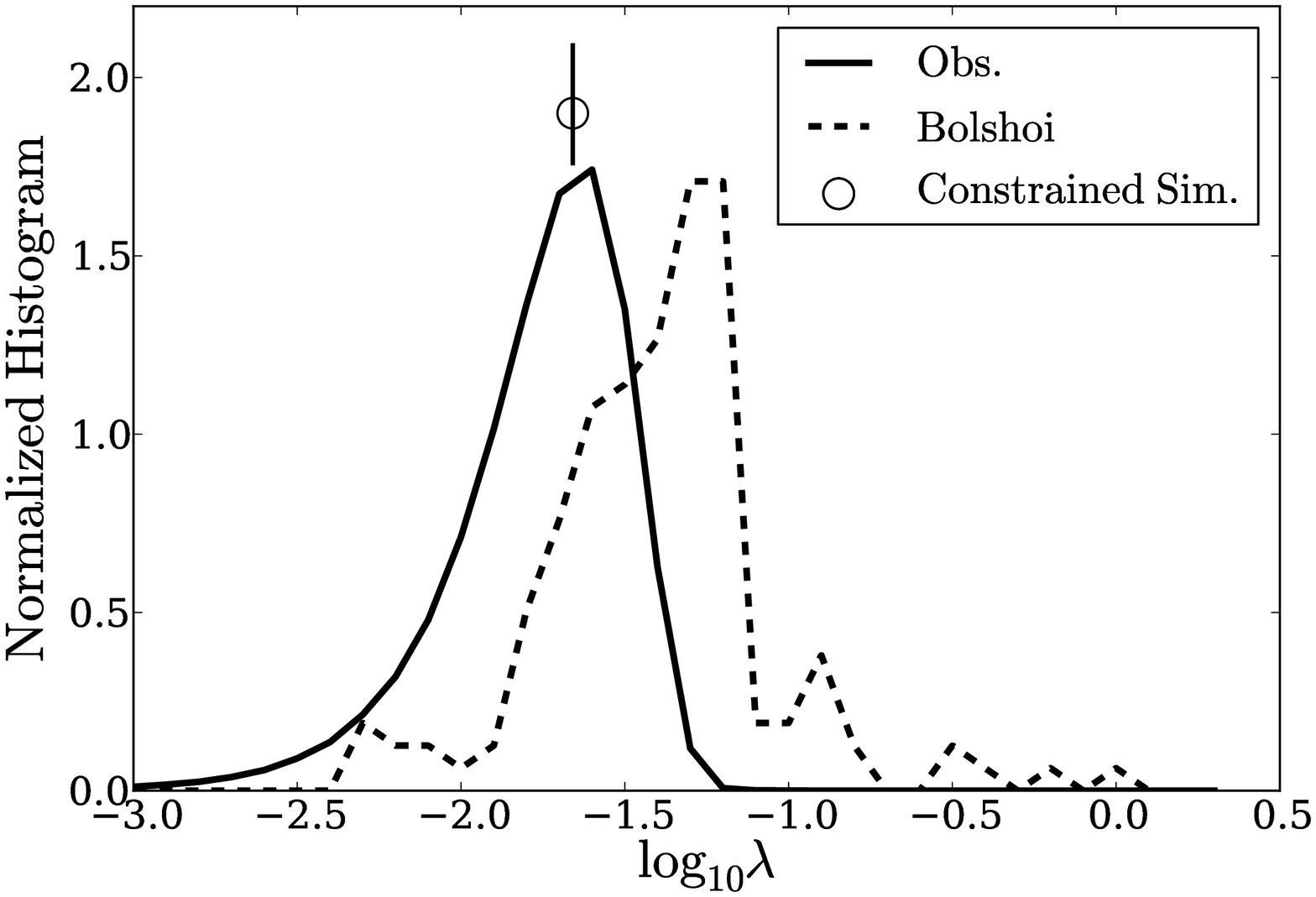}
\caption{{\rm \label{fig:lambda} Normalized histograms of Peebles'
    spin parameter $\lambda$ for the pairs in the Bolshoi simulation
    and its inferred values for the LG from the observational
    constraints from a Monte Carlo simulation. The left (right) panel
    corresponds to the full (reduced) sample. The vertical lines with
    the white dots   represent the values inferred for pairs in the
    constrained 
    simulations.}} 
\label{fig:lambda}
\end{center}
\end{figure*}

\subsection{Dimensionless Spin Parameter}

The dimensionless 
spin parameter $\lambda$ \citep{Peebles1971} is used here to
characterize the dynamical state of the observed and simulated
LGs. The parameter measures the dynamical role of the angular momentum
in terms of the gravitational attraction and is defined by 

\begin{equation}
\lambda = \frac{\mu^{3/2} l_{\rm orb} \sqrt{e_{\rm tot}}}{G M^{5/2}}.
\end{equation}

We compare the distribution for  $\lambda$ obtained from the pair
population in the Bolshoi simulation and the distribution from the  Monte Carlo
simulation used to estimate the uncertainties on $e_{\rm tot}$ and $l_{\rm orb}$. 
Figure \ref{fig:lambda} shows the likelihood distribution of the
observational errors and the spin parameter of the LG-like pairs of
the full (left panel) and the reduced (right panel) samples.  The value
of $\lambda$ of the constrained LGs is shown as well.  

The expected observational value for $\lambda$ is slightly inconsistent with
the statistics derived from halo pairs in $\Lambda$CDM. The spin
parameter of the simulated LG-like objects is skewed towards higher
values compared with the observational estimation.  Inspection
of Figure \ref{fig:lambda} also shows that the $\lambda$ distribution is
close to a log-normal one and thus resembles the distribution of the
spin parameter of DM halos in simulation \citep{Bett2007}.  

The estimated value of $\log_{10}\lambda$ of the
observed LG is $-1.72\pm 0.07$, lower than the mean value of
$\log_{10}\lambda$ of $-1.47\pm0.13$ ($-1.34\pm 0.12$) for the full
(reduced) sample. The values of the spin parameter of the 3
constrained LGs are $\log_{10}(\lambda_{\rm  CLUES})=-2.21$, $-1.72$
and $-1.65$, with the latter value belonging to the sole
constrained LG of the reduced sample.

\section{Conclusions}
We have presented a comparison between the observed kinematics for the
M31 in the galactocentric rest-frame and the expectations for a large
N-body cosmological simulation in the $\Lambda$CDM cosmology. In the
simulation we select a large sample of pairs that fulfill isolation
criteria for the LG. We select a sub-sample which obeys more
stringent observational constraints on the mass and separation of the
MW and M31.

While the observations show that M31 moves towards us  on a highly eccentric orbit,
 the simulation shows that the most common configuration at $z=0$ has
 values $v_{\rm rad, \Lambda CDM}=-80\pm20$\kms and $v_{\rm tan,
   \Lambda CDM}=50\pm10$ \kms, where the error bars are estimated to
 include the results from the two pair samples.

Using the same absolute values for the uncertainty in the observed
velocity components, we find that none of the halos in the reduced sample
are compatible with observations. This makes pairs with the preferred
$\Lambda$CDM values at least $13$ times more common than pairs
compatible with the observational constraint. Additionally, pairs with
a fraction of tangential to radial velocity $f_{\rm t}<0.32$ (similar
to observations) represent $8\%$ to $12\%$ of the pairs.

Approximating the LG as two point masses we express the above
mentioned results in terms of the orbital angular momentum $l_{\rm
  orb}$ and the mechanical energy $e_{\rm tot}$ per unit of reduced
mass. We find that the uncertainties in the tangential velocity, the
square of the norm of the velocity and the total mass in the LG
are less constraining on the number of simulated pairs that are
consistent with the observations. Nevertheless, in the case of the
reduced sample there is a slight tension between simulation and
observation. A reduction by a factor of $2$ in the observational
uncertainty on the radial velocity would clarify this issue. 

We also use the $\lambda$ spin parameter to gauge
the dynamical state of pairs. The values for the orbital angular momentum
and energy, merged into the $\lambda$ spin parameter, are in mild
disagreement with the observational constraints.

In the three pairs from constrained simulations we find kinematics
dominated by radial velocities. However their velocity components
differ from the observational constraints and their mechanical energy
and orbital angular momentum are in broad concordance with
observations. There is only one pair that fulfills all the
separation, total mass constraints and matches the most probable value
for the dimensionless spin parameter $\lambda$ inferred from
observations.

Summarizing, we see a broad agreement in the total angular
momentum and energy and a marked difference with the precise balance
between today's radial and tangential velocities, i.e. head-on
  collisions of MW-like halos, as described by 1-$\sigma$
  uncertainties in observations, are not common in $\Lambda$CDM.

Under the
approximation of conservation of the orbital angular momentum and
mechanical energy, this could only be explained if the initial
conditions for the formation of the Local Group are special in
comparison to the initial conditions of any other pair of dark matter
halos in the $\Lambda$CDM cosmology. Investigating this
  perspective, by a thorough characterization  of the small pair
  sample consistent with observations in terms   of halo properties
  such as concentration, spin and place in the cosmic web
  \citep{Forero-Romero09,Hoffman12} is  underway (Bustamante et al. in
  prep.).    

The results presented in this Letter open a new window into the
  question of how unique, if at all, is the LG in a cosmological
  context. This will continue to be studied within the framework of
  the CLUES project.

\label{sec:conclusions}
\section*{Acknowledgments}  
JEF-R acknowledges financial support from Universidad de los Andes through its {\it Fondo de
  Apoyo a Profesores Asistentes} and the Peter and Patricia Gruber
Foundation through its fellowship administered by the IAU. JEF-R also acknowledges early discussions with
Alejandro Garc\'ia that motivated this work. YH has been supported by
the ISF (1013/12) 

GY  acknowledges support from MINECO through research grants AYA 2009
13875 -C03-02 , AYA 2012 31101  and Consolider Syec CSD 2007 0050.
Support from Comunidad. De Madrid  through ASTROMADRID research grant
is also acknowledge  

The data and source code and instructions to replicate the results of
this paper can be found here {\texttt{https://github.com/forero/LG\_Kinematics/}}. Thanks to the IPython
community \citep{IPython}. Thanks to Jessica Kirkpatrick for releasing
her Python code to make nice plots of 2D histograms.  

The MultiDark Database used in this paper  was constructed as part of
the activities of the German Astrophysical Virtual Observatory as
result of a collaboration between the Leibniz-Institute for
Astrophysics Potsdam (AIP) and the Spanish MultiDark Consolider
Project CSD2009-00064. The Bolshoi simulation was run on the NASA's
Pleiades supercomputer at the NASA Ames Research Center. The CLUES simulation
was run at LRZ Munich.

\bibliographystyle{apj}

\begin{thebibliography}{30}
\expandafter\ifx\csname natexlab\endcsname\relax\def\natexlab#1{#1}\fi

\bibitem[{{Besla} {et~al.}(2007){Besla}, {Kallivayalil}, {Hernquist},
  {Robertson}, {Cox}, {van der Marel}, \& {Alcock}}]{Besla2007}
{Besla}, G., {Kallivayalil}, N., {Hernquist}, L., {Robertson}, B., {Cox},
  T.~J., {van der Marel}, R.~P., \& {Alcock}, C. 2007, \apj, 668, 949

\bibitem[{{Bett} {et~al.}(2007){Bett}, {Eke}, {Frenk}, {Jenkins}, {Helly}, \&
  {Navarro}}]{Bett2007}
{Bett}, P., {Eke}, V., {Frenk}, C.~S., {Jenkins}, A., {Helly}, J., \&
  {Navarro}, J. 2007, \mnras, 376, 215

\bibitem[{{Busha} {et~al.}(2011){Busha}, {Marshall}, {Wechsler}, {Klypin}, \&
  {Primack}}]{Busha2011}
{Busha}, M.~T., {Marshall}, P.~J., {Wechsler}, R.~H., {Klypin}, A., \&
  {Primack}, J. 2011, \apj, 743, 40

\bibitem[{{Forero-Romero} {et~al.}(2009){Forero-Romero}, {Hoffman},
  {Gottl{\"o}ber}, {Klypin}, \& {Yepes}}]{Forero-Romero09}
{Forero-Romero}, J.~E., {Hoffman}, Y., {Gottl{\"o}ber}, S., {Klypin}, A., \&
  {Yepes}, G. 2009, \mnras, 396, 1815

\bibitem[{{Forero-Romero} {et~al.}(2011){Forero-Romero}, {Hoffman}, {Yepes},
  {Gottl{\"o}ber}, {Piontek}, {Klypin}, \& {Steinmetz}}]{ForeroRomero2011}
{Forero-Romero}, J.~E., {Hoffman}, Y., {Yepes}, G., {Gottl{\"o}ber}, S.,
  {Piontek}, R., {Klypin}, A., \& {Steinmetz}, M. 2011, \mnras, 417, 1434

\bibitem[{{Fouquet} {et~al.}(2012){Fouquet}, {Hammer}, {Yang}, {Puech}, \&
  {Flores}}]{Fouquet2012}
{Fouquet}, S., {Hammer}, F., {Yang}, Y., {Puech}, M., \& {Flores}, H. 2012,
  \mnras, 427, 1769

\bibitem[{{Freeman} \& {Bland-Hawthorn}(2002)}]{Freeman2002}
{Freeman}, K., \& {Bland-Hawthorn}, J. 2002, \araa, 40, 487

\bibitem[{{Gottl{\"o}ber} {et~al.}(2010){Gottl{\"o}ber}, {Hoffman}, \&
  {Yepes}}]{clues2010}
{Gottl{\"o}ber}, S., {Hoffman}, Y., \& {Yepes}, G. 2010, ArXiv e-prints

\bibitem[{{Hoffman} {et~al.}(2012){Hoffman}, {Metuki}, {Yepes},
  {Gottl{\"o}ber}, {Forero-Romero}, {Libeskind}, \& {Knebe}}]{Hoffman12}
{Hoffman}, Y., {Metuki}, O., {Yepes}, G., {Gottl{\"o}ber}, S., {Forero-Romero},
  J.~E., {Libeskind}, N.~I., \& {Knebe}, A. 2012, \mnras, 425, 2049

\bibitem[{{Jarosik} {et~al.}(2011){Jarosik}, {Bennett}, {Dunkley}, {Gold},
  {Greason}, {Halpern}, {Hill}, {Hinshaw}, {Kogut}, {Komatsu}, {Larson},
  {Limon}, {Meyer}, {Nolta}, {Odegard}, {Page}, {Smith}, {Spergel}, {Tucker},
  {Weiland}, {Wollack}, \& {Wright}}]{Jarosik2011}
{Jarosik}, N., {Bennett}, C.~L., {Dunkley}, J., {Gold}, B., {Greason}, M.~R.,
  {Halpern}, M., {Hill}, R.~S., {Hinshaw}, G., {Kogut}, A., {Komatsu}, E.,
  {Larson}, D., {Limon}, M., {Meyer}, S.~S., {Nolta}, M.~R., {Odegard}, N.,
  {Page}, L., {Smith}, K.~M., {Spergel}, D.~N., {Tucker}, G.~S., {Weiland},
  J.~L., {Wollack}, E., \& {Wright}, E.~L. 2011, \apjs, 192, 14

\bibitem[{{Karachentsev} {et~al.}(2004){Karachentsev}, {Karachentseva},
  {Huchtmeier}, \& {Makarov}}]{Karachentsev04}
{Karachentsev}, I.~D., {Karachentseva}, V.~E., {Huchtmeier}, W.~K., \&
  {Makarov}, D.~I. 2004, \aj, 127, 2031

\bibitem[{{Kazantzidis} {et~al.}(2008){Kazantzidis}, {Bullock}, {Zentner},
  {Kravtsov}, \& {Moustakas}}]{Kazantzidis2008}
{Kazantzidis}, S., {Bullock}, J.~S., {Zentner}, A.~R., {Kravtsov}, A.~V., \&
  {Moustakas}, L.~A. 2008, \apj, 688, 254

\bibitem[{{Klypin} {et~al.}(1999{\natexlab{a}}){Klypin}, {Gottl{\"o}ber},
  {Kravtsov}, \& {Khokhlov}}]{KlypinBDM}
{Klypin}, A., {Gottl{\"o}ber}, S., {Kravtsov}, A.~V., \& {Khokhlov}, A.~M.
  1999{\natexlab{a}}, \apj, 516, 530

\bibitem[{{Klypin} {et~al.}(1999{\natexlab{b}}){Klypin}, {Kravtsov},
  {Valenzuela}, \& {Prada}}]{Klypin99}
{Klypin}, A., {Kravtsov}, A.~V., {Valenzuela}, O., \& {Prada}, F.
  1999{\natexlab{b}}, \apj, 522, 82

\bibitem[{{Klypin} {et~al.}(2011){Klypin}, {Trujillo-Gomez}, \&
  {Primack}}]{Bolshoi}
{Klypin}, A.~A., {Trujillo-Gomez}, S., \& {Primack}, J. 2011, \apj, 740, 102

\bibitem[{{Knebe} {et~al.}(2011){Knebe}, {Libeskind}, {Doumler}, {Yepes},
  {Gottl{\"o}ber}, \& {Hoffman}}]{Knebe2011}
{Knebe}, A., {Libeskind}, N.~I., {Doumler}, T., {Yepes}, G., {Gottl{\"o}ber},
  S., \& {Hoffman}, Y. 2011, \mnras, 417, L56

\bibitem[{{Liu} {et~al.}(2011){Liu}, {Gerke}, {Wechsler}, {Behroozi}, \&
  {Busha}}]{Liu2011}
{Liu}, L., {Gerke}, B.~F., {Wechsler}, R.~H., {Behroozi}, P.~S., \& {Busha},
  M.~T. 2011, \apj, 733, 62

\bibitem[{{McConnachie} {et~al.}(2009){McConnachie}, {Irwin}, {Ibata},
  {Dubinski}, {Widrow}, {Martin}, {C{\^o}t{\'e}}, {Dotter}, {Navarro},
  {Ferguson}, {Puzia}, {Lewis}, {Babul}, {Barmby}, {Bienaym{\'e}}, {Chapman},
  {Cockcroft}, {Collins}, {Fardal}, {Harris}, {Huxor}, {Mackey},
  {Pe{\~n}arrubia}, {Rich}, {Richer}, {Siebert}, {Tanvir}, {Valls-Gabaud}, \&
  {Venn}}]{pandas09}
{McConnachie}, A.~W., {Irwin}, M.~J., {Ibata}, R.~A., {Dubinski}, J., {Widrow},
  L.~M., {Martin}, N.~F., {C{\^o}t{\'e}}, P., {Dotter}, A.~L., {Navarro},
  J.~F., {Ferguson}, A.~M.~N., {Puzia}, T.~H., {Lewis}, G.~F., {Babul}, A.,
  {Barmby}, P., {Bienaym{\'e}}, O., {Chapman}, S.~C., {Cockcroft}, R.,
  {Collins}, M.~L.~M., {Fardal}, M.~A., {Harris}, W.~E., {Huxor}, A., {Mackey},
  A.~D., {Pe{\~n}arrubia}, J., {Rich}, R.~M., {Richer}, H.~B., {Siebert}, A.,
  {Tanvir}, N., {Valls-Gabaud}, D., \& {Venn}, K.~A. 2009, \nat, 461, 66

\bibitem[{{Moore} {et~al.}(1999){Moore}, {Ghigna}, {Governato}, {Lake},
  {Quinn}, {Stadel}, \& {Tozzi}}]{Moore99}
{Moore}, B., {Ghigna}, S., {Governato}, F., {Lake}, G., {Quinn}, T., {Stadel},
  J., \& {Tozzi}, P. 1999, \apjl, 524, L19

\bibitem[{{Peebles}(1971)}]{Peebles1971}
{Peebles}, P.~J.~E. 1971, \aap, 11, 377

\bibitem[{{Peebles} \& {Nusser}(2010)}]{Peebles2010}
{Peebles}, P.~J.~E., \& {Nusser}, A. 2010, \nat, 465, 565

\bibitem[{P\'erez \& Granger(2007)}]{IPython}
P\'erez, F., \& Granger, B.~E. 2007, {C}omput. {S}ci. {E}ng., 9, 21

\bibitem[{{Purcell} \& {Zentner}(2012)}]{Purcell2012}
{Purcell}, C.~W., \& {Zentner}, A.~R. 2012, {JCAP}, 12, 7

\bibitem[{{Ribas} {et~al.}(2005){Ribas}, {Jordi}, {Vilardell}, {Fitzpatrick},
  {Hilditch}, \& {Guinan}}]{ribas05}
{Ribas}, I., {Jordi}, C., {Vilardell}, F., {Fitzpatrick}, E.~L., {Hilditch},
  R.~W., \& {Guinan}, E.~F. 2005, \apjl, 635, L37

\bibitem[{{Riebe} {et~al.}(2011){Riebe}, {Partl}, {Enke}, {Forero-Romero},
  {Gottloeber}, {Klypin}, {Lemson}, {Prada}, {Primack}, {Steinmetz}, \&
  {Turchaninov}}]{2011arXiv1109.0003R}
{Riebe}, K., {Partl}, A.~M., {Enke}, H., {Forero-Romero}, J., {Gottloeber}, S.,
  {Klypin}, A., {Lemson}, G., {Prada}, F., {Primack}, J.~R., {Steinmetz}, M.,
  \& {Turchaninov}, V. 2011, ArXiv e-prints

\bibitem[{{Teyssier} {et~al.}(2012){Teyssier}, {Johnston}, \&
  {Kuhlen}}]{Teyssier2012}
{Teyssier}, M., {Johnston}, K.~V., \& {Kuhlen}, M. 2012, \mnras, 426, 1808

\bibitem[{{Tikhonov} \& {Klypin}(2009)}]{Anton09}
{Tikhonov}, A.~V., \& {Klypin}, A. 2009, \mnras, 395, 1915

\bibitem[{{Tollerud} {et~al.}(2011){Tollerud}, {Boylan-Kolchin}, {Barton},
  {Bullock}, \& {Trinh}}]{Tollerud2011}
{Tollerud}, E.~J., {Boylan-Kolchin}, M., {Barton}, E.~J., {Bullock}, J.~S., \&
  {Trinh}, C.~Q. 2011, \apj, 738, 102

\bibitem[{{van der Marel} {et~al.}(2012){van der Marel}, {Fardal}, {Besla},
  {Beaton}, {Sohn}, {Anderson}, {Brown}, \& {Guhathakurta}}]{vanderMarel12}
{van der Marel}, R.~P., {Fardal}, M., {Besla}, G., {Beaton}, R.~L., {Sohn},
  S.~T., {Anderson}, J., {Brown}, T., \& {Guhathakurta}, P. 2012, \apj, 753, 8

\bibitem[{{van der Marel} \& {Guhathakurta}(2008)}]{vanderMarel08}
{van der Marel}, R.~P., \& {Guhathakurta}, P. 2008, \apj, 678, 187

\end{thebibliography}

\end{document}